\documentclass[aps,preprint,superscriptaddress,showpacs,floatfix]{revtex4}%
\usepackage{amsfonts}
\usepackage{amsmath}
\usepackage{amssymb}
\usepackage{graphicx}
\usepackage{graphicx}%
\begin{document}
\title{Application of M\"{o}ssbauer effect to study subnanometer harmonic
displacements in a thin solid}
\author{R. N. Shakhmuratov}
\affiliation{Kazan Physical-Technical Institute, Russian Academy of Sciences,
10/7 Sibirsky Trakt, Kazan 420029 Russia}
\affiliation{Kazan Federal University, 18 Kremlyovskaya Street, Kazan 420008 Russia}
\author{F. G. Vagizov}
\affiliation{Kazan Federal University, 18 Kremlyovskaya Street, Kazan 420008 Russia}
\pacs{42.25.Bs, 42.50.Gy, 42.50.Nn}
\date{{ \today}}

\begin{abstract}
We measure subnanometer displacements of thin samples vibrated by
piezotransducer. Samples contain $^{57}$Fe nuclei, which are exposed to 14.4
keV $\gamma$-radiation. Vibration produces sidebands from a single absorption
line of the sample. The sideband intensities depend on the vibration amplitude
and its distribution along the sample. We developed a model of this
distribution, which adequately describes the spectra of powder and stainless
steel (SS) absorbers. We propose to filter $\gamma$-radiation through a small
hole in the mask, placed before the absorber. In this case only a small spot
of the vibrated absorber is observed. We found for SS foil that nuclei,
exposed to $\gamma$-radiation in this small spot, vibrate with almost the same
amplitudes whose difference does not exceed a few picometers.

\end{abstract}
\maketitle

\section{Introduction}

A scanning tunneling microscope (STM) demonstrates remarkable lateral and
depth resolution. STM proved to be an excellent instrument for imaging
surfaces at the atomic level. Its development earned its inventors, Binning
and Rohrer, the Nobel prize in Physics in 1986. Information is acquired by
monitoring the tunneling current as a function of tip position, applied
voltage, and local density of states. If local density of states is not well
known, then STM needs calibration. To solve the problem of absolute
calibration of the conducting tip displacement an etalon step is usually grown
on the surface of the sample. Recently, fluorescence microscopy with nanoscale
spatial resolution was invented, see the review by Stefan Hell Ref.
\cite{Hell}, the Nobel prize winner in Chemistry in 2014. This method allows
to overcome diffraction limit. Meanwhile, if we could apply $\gamma$-radiation
for spatial measurements, then Angstrom resolution scale could be achieved.

In this paper we discuss the application of $\gamma$-photons for sub-nanometer
spatial measurements with sub-Angstrom resolution. We employ 14.4 keV $\gamma
$-photons, emitted by radioactive $^{57}$Co with a wavelength 86 pm. Since we
are not able to focus $\gamma$-radiation field and direct it to a desirable
spot on the sample, we address to the spectral measurements. In our case, not
a fluorescence, but transmission spectra of the sample, containing resonant
nuclei $^{57}$Fe, are proposed to be measured. We expect high depth
resolution, while lateral resolution could be controlled by a mask with a
small hole, which is moved along the surface of the sample. We expect that our
method could provide controllable and calibrated displacements of the surface
with sub-Angstrom resolution for calibrating, for example, STM.

In our method the sample containing resonant nuclei is mechanically vibrated.
As a result, along with the main absorption line, a system of satellites
appears in the spectrum, spaced apart at distances that are multiples of the
vibration frequency. Line intensities of this comb structure are very
sensitive to the vibration amplitude, which is extremely small (Angstrom or
even smaller). For the stainless steel foil, which is illuminated by $\gamma
$-radiation transmitted through a small hole in the mask placed before the
absorber, we measured displacements of the order of dozens of picometers with
the accuracy of a few picometers.

The influence of extremely-small-amplitude mechanical vibrations of the
absorbers containing M\"{o}ssbauer nuclei attracted attention since the
invention of M\"{o}ssbauer spectroscopy. M\"{o}ssbauer sidebands, produced
from a single parent line by absorber vibration, were observed in many
different samples
\cite{Ruby,Cranshaw,Mishroy,Kornfeld,Walker,Mkrtchyan77,Perlow,Monahan,
Mkrtchyan79,Shvydko89,Shvydko92}. However, the intensity of the sidebands has
not been yet satisfactory explained \cite{Walker,Mkrtchyan77,Mkrtchyan79}.
There are two models of coherent and incoherent vibrations of nuclei in the
absorber \cite{Cranshaw,Walker, Mishroy,Kornfeld,Mkrtchyan77,Mkrtchyan79}.
Coherent model implies piston-like vibration of the absorber with frequency
$\Omega$ and phase $\psi$ along the propagation direction of $\gamma$ quanta.
This model predicts the intensity of the $n$-th sideband proportional to the
square of Bessel function $J_{n}^{2}(m)$, where $m=2\pi a_{0}/\lambda$ is the
modulation index, which is proportional to the ratio of the amplitude of the
harmonic displacements $a_{z}(t)=a_{0}\sin(\Omega t+\psi)$ and the wavelenght
$\lambda$ of $\gamma$-photon. Incoherent model, proposed by Abraham
\cite{Abragam}, is based on the Rayleigh distribution of the nuclear-vibration
amplitudes in the absorber predicting the intensity proportional to
$\exp(-m^{2})I_{n}(m^{2})$, where $I_{n}(m^{2})$ is the modified Bessel
function. However, both models or their combinations cannot describe perfectly
all absorption spectra, which are experimentally observed for samples of
different mechanical properties and chemical composition. To support this
statement we refer to Chien and Walker who pointed out in Ref. \cite{Walker}
that while unequivocally measured intensities agree qualitatively with
Abragam's sideband theory, no existing theory at present can account
quantitatively for the sideband intensities since the amplitude distribution
that satisfactorily describes the data are not known at present.

We propose a heuristic distribution of nuclear-vibration amplitudes, which is
derived from the Gaussian distribution with appropriate modifications. Our
model provides good fitting of experimental spectra. Depending on a parameter
of the model $\sigma$, the proposed distribution tends to a delta-like,
inherent to the coherent model if $\sigma\rightarrow0$, or it tends to the
Rayleigh distribution inherent to the incoherent model if $\sigma
\rightarrow0.72$.

The proposed model allows to determine from experimental data the amplitude of
subnanometer harmonic displacements of the absorber with an accuracy less than
half Angstrom. Our method consists of two steps. First, we apply our heuristic
distribution to fit experimental spectra to the model. This fitting gives the
parameter $\sigma$, which specifies the appropriate distribution of the
displacements irrespective to their location in the sample. Second, we
construct an actual distribution of the vibration amplitudes across the
surface of the absorber, which is consistent with our heuristic distribution.

Two absorbers are experimentally studied, i.e., K$_{4}$Fe(CN)$_{6}\cdot
3$H$_{2}$O powder enriched by $^{57}$Fe and stainless steel (SS) foil with
natural abundance of $^{57}$Fe. For powder, the distribution of the
powder-grain displacements, obtained from the spectrum fitting, is close to
the continuous uniform distribution with wide scattering of the vibration
amplitudes, which is very different from the Rayleigh distribution. For SS foil
we found that the displacement distribution along the surface of a thin foil is
close to the narrow bell-shape distribution. Physical interpretation of these
results is discussed.

For SS foil we experimentally verified our conclusions placing a mask just in
the front of the absorber. We made a hole in the mask and compared the
observed M\"{o}ssbauer spectra with our theoretical predictions. We observed a
change of M\"{o}ssbauer spectra with decrease of the size of the hole, which
firmly supports our model. We moved the narrowest hole of the mask along the
surface of the vibrated SS foil and could detect the change of the vibration
amplitude along the SS foil, which is deduced from the spectrum analysis. In
addition to a scientific value of our model giving an explanation of
properties of the M\"{o}ssbauer sidebands produced by the absorber vibration,
we expect that our findings could give an impetus to the development of the
method measuring extra-small displacements with an accuracy less than half Angstrom.

\section{Coherent and incoherent models of the mechanical vibrations of the
absorber}

The propagation of $\gamma$ radiation through a resonant M\"{o}ssbauer medium
vibrating with frequency $\Omega$ may be treated classically \cite{Lynch}. In
this approach the radiation field from the source nucleus after passing
through a small diaphragm is approximated as a plane wave propagating along
the direction $\mathbf{z}$. In the coordinate system rigidly bounded to the
absorbing sample, the field, seen by the absorber nuclei, is described by
\begin{equation}
E_{S}(t-t_{0})\propto\theta(t-t_{0})e^{-(i\omega_{S}+\Gamma_{0}/2)(t-t_{0}%
)+ikz+i\varphi(t)}, \label{Eq1}%
\end{equation}
where $\omega_{S}$ and $k$ are the carrier frequency and the wave number of
the radiation field, $1/\Gamma_{0}$ is the lifetime of the excited state of
the emitting source nucleus, $t_{0}$ is the instant of time when the excited
state is formed, $\Theta(t-t_{0})$ is the Heaviside step-function,
$\varphi(t)=2\pi a_{z}(t)/\lambda=m\sin(\Omega t+\psi)$ is a time dependent
phase of the field due to a piston-like periodical displacement $a_{z}(t)$ of
the absorber with respect to the source, $\psi$ is a vibration phase, and
$\lambda$ is the radiation wavelength.

The radiation field (\ref{Eq1}) can be expressed as polychromatic radiation
with a set of spectral lines $\omega_{S} - n\Omega$ ($n=0$, $\pm 1$, $\pm2$,
...), i.e.,
\begin{equation}
E_{S}(t-t_{0})=E_{C}(t-t_{0})e^{-i\omega_{S}(t-t_{0})+ikz}\sum_{n=-\infty
}^{+\infty}J_{n}(m)e^{in(\Omega t+\psi)}, \label{Eq2}%
\end{equation}
where $E_{C}(t-t_{0})=E_{0}\theta(t-t_{0})e^{-\Gamma_{0}(t-t_{0})/2}$ is the
common part of the field components, $E_{0}$ is the field amplitude, and
$J_{n}(a)$ is the Bessel function of the $n$th order. The Fourier transform
of this field has a frequency comb structure
\begin{equation}
E_{S}(\omega)=E_{0}\sum_{n=-\infty}^{+\infty}\frac{J_{n}(a)e^{in(\Omega
t_{0}+\psi)}}{\Gamma_{0}/2+i(\omega_{S}-n\Omega-\omega)}, \label{Eq3}%
\end{equation}
where for shortening of notations the exponential factor with $ikz$ is
omitted. From this expression, it follows that the vibrating absorber `sees'
the incident radiation as an equidistant frequency comb with spectral
components $\omega_{S}-n\Omega$ whose amplitudes are proportional to the
Bessel function $J_{n}(m)$.

The Fourier transform of the radiation field is changed at the exit of the
resonant absorber as (see \cite{Lynch,Monahan,Vagizov,Shakhmuratov15})
\begin{equation}
E_{\text{out}}(\omega)=E_{0}\sum_{n=-\infty}^{+\infty}\frac{J_{n}%
(m)\exp\left[  in(\Omega t_{0}+\psi)-\frac{b}{\Gamma_{A}/2+i(\omega_{A}%
-\omega)}\right]  }{\Gamma_{0}/2+i(\omega_{S}-n\Omega-\omega)}, \label{Eq4}%
\end{equation}
where $\omega_{A}$ and $\Gamma_{A}$ are the frequency and linewidth of the
absorber, $b=T_{A}\Gamma_{0}/4$ is the parameter depending on the effective
thickness of the absorber $T_{A}=fn_{A}\sigma_{A}$, $f$ is the Debye-Waller
factor, $n_{A}$ is the number of $^{57}$Fe nuclei per unit area of the
absorber, and $\sigma_{A}$ is the resonance absorption cross section. The
source linewidth $\Gamma_{S}$ can be different from $\Gamma_{0}$ due to the
contribution of the environment of the emitting nucleus in the source. In this
case $\Gamma_{0}$ can be simply substituted by $\Gamma_{S}$ in Eq.
(\ref{Eq4}). Here, nonresonant absorption is disregarded. Recoil processes in
nuclear absorption and emission are not taken into account assuming that
recoilless fraction (Debye-Waller factor) is $f=1$. These processes can be
easily taken into account in experimental data analysis.

Time dependence of the amplitude of the output radiation field is found by
inverse Fourier transformation%
\begin{equation}
E_{\text{out}}(t-t_{0})=\frac{1}{2\pi}\int_{-\infty}^{+\infty}E_{\text{out}%
}(\omega)e^{-i\omega(t-t_{0})}d\omega. \label{Eq5}%
\end{equation}
In the laboratory reference frame this field is transformed as%
\begin{equation}
E_{\text{lab}}(t-t_{0})=E_{\text{out}}(t-t_{0})e^{-i\varphi(t)}. \label{Eq6}%
\end{equation}
Since the fields $E_{\text{lab}}(t-t_{0})$ and $E_{\text{out}}(t-t_{0})$
differ only in the phase $\varphi(t)$, the intensity seen by the detector,
$I_{\text{lab}}(t-t_{0})=\left\vert E_{\text{lab}}(t-t_{0})\right\vert ^{2}$,
coincides with the intensity of the radiation field in the vibrated reference
frame $I_{\text{lab}}(t-t_{0})=\left\vert E_{\text{lab}}(t-t_{0})\right\vert
^{2}$, i.e.,%
\begin{equation}
I_{\text{lab}}(t-t_{0})=I_{\text{out}}(t-t_{0}). \label{Eq7}%
\end{equation}
This condition is valid if we use the detection scheme, which is not sensitive
to the spectral content of the radiation field filtered by the vibrated
absorber. If the second absorber (spectrum analyzer) is placed between the
vibrated source and detector, then another description of the radiation
intensity is necessary \cite{Shvydko89,Shvydko92}.

Since we don't use a second single line resonance filter analyzing the spectra
of $\gamma$ radiation emerging from the vibrated absorber, the intensity of
the field, registered by a detector, can be described by expression%
\begin{equation}
I_{\text{out}}(t-t_{0})=\frac{1}{(2\pi)^{2}}\int_{-\infty}^{+\infty}%
d\omega_{1}\int_{-\infty}^{+\infty}d\omega_{2}E_{\text{out}}(\omega
_{1})E_{\text{out}}^{\ast}(\omega_{2})e^{i(\omega_{2}-\omega_{1})(t-t_{0})}.
\label{Eq8}%
\end{equation}
Thus, in our case the radiation intensity at the exit of the vibrating
absorber is the same if the source is vibrated instead of absorber.

Frequency-domain M\"{o}ssbauer spectrum is measured by counting the number of
photons, detected within long time windows of the same duration for all
resonant detunings, which are varied by changing the value of a velocity of
the M\"{o}ssbauer drive moving the source. Time windows are not
synchronized with the mechanical vibration and their duration $T_{\text{w}}$
is much longer than the oscillation period $T_{\text{osc}}=2\pi/\Omega$. Since
the emission time of $\gamma$-photons is random, the observed radiation
intensity is averaged over $t_{0}$
\begin{equation}
\left\langle I_{\text{out}}(t-t_{0})\right\rangle _{t_{0}}\propto\frac
{1}{T_{w}}\int_{-T_{\text{w}}/2}^{+T_{\text{w}}/2}I_{\text{out}}%
(t-t_{0})dt_{0}, \label{Eq9}%
\end{equation}
where for simplicity we assume that $T_{\text{w}}\rightarrow\infty$. Then the
observed number of photon counts, which is proportional to the intensity,
i.e., $N_{\text{out}}(\Delta)=\left\langle I_{\text{out}}(t-t_{0}%
)\right\rangle _{t_{0}}$, varies with the change of the resonant detuning
$\Delta=\omega_{A}-\omega_{S}$ as%
\begin{equation}
N_{\text{out}}(\Delta)=\sum_{n=-\infty}^{+\infty}J_{n}^{2}(m)B_{n}(\Delta),
\label{Eq10}%
\end{equation}
where%
\begin{equation}
B_{n}(\Delta)=\frac{\Gamma_{S}}{2\pi}\int_{-\infty}^{+\infty}\frac
{e^{-\frac{b\Gamma_{A}}{(\Gamma_{A}/2)^{2}+(\Delta+n\Omega-\omega)^{2}}}%
}{(\Gamma_{S}/2)^{2}+\omega^{2}}d\omega. \label{Eq11}%
\end{equation}

\subsection{Coherent model}

\begin{figure}[ptb]
\resizebox{0.4\textwidth}{!}{\includegraphics{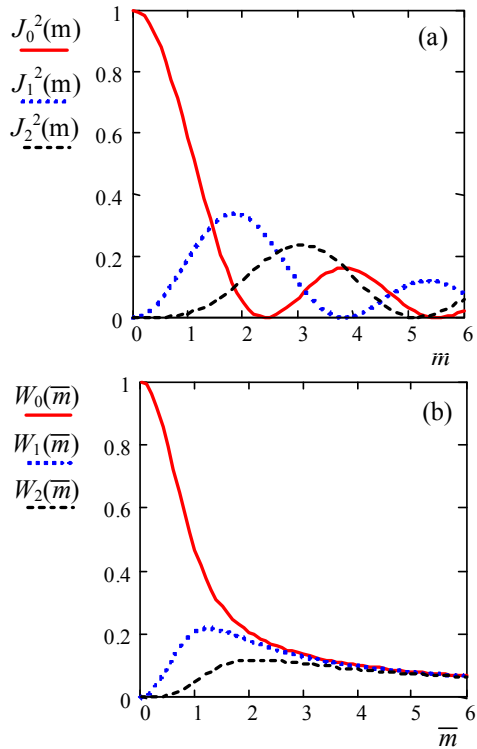}}\caption{(color on
line) (a) The dependence of the intensities of the central component (solid
line in red), first satellite (dotted line in blue), and second satellite
(dashed line in black) on the modulation index $m$ for the coherent model. (b)
The dependence of the averaged intensities of the central component (solid
line in red), first satellite (dotted line in blue), and second satellite
(dashed line in black) on $\bar{m}$, which is the square root of the
modulation index deviation $\langle m^{2}\rangle$ in the incoherent model.}%
\label{fig:1}%
\end{figure}
If all the nuclei in the absorber vibrate with the same phase and
amplitude, then a single parent line is transformed into a set of spectral lines
$\omega_{S}\pm n\Omega$ ($n=0,1,2,...$) spaced apart at distances that are
multiples of the oscillation frequency. The intensity of the $n$th sideband is
given by the square of the Bessel function $J_{n}^{2}(m)$. According to this
theoretical prediction the line intensities oscillate with increase of the
modulation index $m$, see Fig. 1(a). For example, the first sidebands, whose
intensities are proportional to $J_{\pm1}^{2}(m)$ take their global maxima
when $m=1.8$, while the intensity of the central component $J_{0}^{2}(m)$ is
zero if $m=2.4$. A model of uniform and phased vibrations of all the nuclei in
the absorber is named the coherent model. Unfortunately, experiments with powder
absorbers did not demonstrate oscillatory dependence of the sideband intensities
on the modulation index. Even in many experiments the intensity of the central
component was always the strongest, while the intensity of the satellites,
initially increasing with the modulation amplitude increase, then
monotonically decreased as a function of $m$. Meanwhile, experiments with
stainless foil \cite{Mkrtchyan77,Mkrtchyan79} showed appreciable decrease of
the central component of the spectrum to the level of the sidebands with increase
of the modulation index $m$ and some oscillating dependencies of the spectral
components on $m$.

\subsection{Incoherent model}

The incoherent model was proposed \cite{Ruby,Cranshaw,
Mishroy,Kornfeld,Walker} to explain the discrepancy between the coherent model
and the experiment. It is based on the Abragam model \cite{Abragam} suggesting
that the motion of nuclei in the absorber can be described as the sum of two
vibrations along the propagation direction $\mathbf{z}$ of $\gamma$ photon,
i.e., $\mathbf{a}_{z}(t)=a_{x}\cos(\Omega t+\psi)+a_{y}\sin(\Omega t+\psi)$,
where $a_{x}$ and $a_{y}$ are the amplitudes, which are different for
different coordinates $x$ and $y$ in the plane of the absorber surface and in
this sense $\mathbf{a}_{z}(t)$ is a vector \cite{Cranshaw}. These amplitudes
are Gaussian-distributed, centered at zero, and independent. Then, their
distribution is described by the function%
\begin{equation}
G(a_{x,y},\left\langle a_{x,y}^{2}\right\rangle )=\frac{\exp\left(
-\frac{a_{x,y}^{2}}{2\left\langle a_{x,y}^{2}\right\rangle }\right)  }%
{\sqrt{2\pi\left\langle a_{x,y}^{2}\right\rangle }}, \label{Eq12}%
\end{equation}
where $\left\langle a_{x,y}^{2}\right\rangle $ is variance, which is not zero,
while $\left\langle a_{x,y}\right\rangle =0$. The amplitude of the vector
$\mathbf{a}_{z}(t)$ is $a_{z}=\sqrt{a_{x}^{2}+a_{y}^{2}}$. If $\left\langle
a_{x}^{2}\right\rangle =\left\langle a_{y}^{2}\right\rangle =\overline{a^{2}}%
$, the amplitude $a_{z}$ is distributed as%
\begin{equation}
P_{R}(a_{z},\overline{a^{2}})=\int_{-\infty}^{+\infty}da_{x}\int_{-\infty
}^{+\infty}da_{y}\frac{\exp\left(  -\frac{a_{x}^{2}+a_{y}^{2}}{2\overline
{a^{2}}}\right)  }{2\pi\overline{a^{2}}}\delta(a_{z}-\sqrt{a_{x}^{2}+a_{y}%
^{2}}). \label{Eq13}%
\end{equation}
In a polar coordinate system $(r,\phi)$ this distribution is transformed to%
\begin{equation}
P_{R}(a_{z},\overline{a^{2}})=\int_{0}^{2\pi}d\phi\int_{-\infty}^{+\infty
}dr\frac{\exp\left(  -\frac{r^{2}}{2\overline{a^{2}}}\right)  }{2\pi
\overline{a^{2}}}\delta(r-a_{z}), \label{Eq14}%
\end{equation}
which gives the Rayleigh distribution%
\begin{equation}
P_{R}(a_{z},\overline{a^{2}})=\frac{a_{z}}{\overline{a^{2}}}\exp\left(
-\frac{a_{z}^{2}}{2\overline{a^{2}}}\right)  . \label{Eq15}%
\end{equation}
Averaging the intensity of the $n$th sideband with this distribution%
\begin{equation}
W_{n}(\overline{m})=\int_{0}^{\infty}P_{R}(a_{z},\overline{a^{2}})J_{n}\left(
\frac{2\pi a_{z}}{\lambda}\right)  da_{z}, \label{Eq16}%
\end{equation}
one obtains%
\begin{equation}
W_{n}(\overline{m})=e^{-\left\langle m^{2}\right\rangle }I_{n}\left(
\left\langle m^{2}\right\rangle \right)  , \label{Eq17}%
\end{equation}
where $\left\langle m^{2}\right\rangle =(2\pi/\lambda)^{2}\overline{a^{2}}$
and $\overline{m}=\sqrt{\left\langle m^{2}\right\rangle }$. The dependence of
the components $W_{n}(\overline{m})$ on $\overline{m}$ are shown in Fig. 1(b)
for $n=0$, $1$, and $2$ .

The physical difference between coherent and incoherent models is well
formulated, for example, in Ref. \cite{Gupta} where two cases of the nuclear
vibrations are distinguished. If the relaxation time of the generated phonons,
$\tau_{ph}$, is much greater than the lifetime of the excited nucleus,
$1/\Gamma_{0}$, then the acoustic wave can be considered as coherent and the
coherent model is applicable. If $\tau_{ph}\ll1/\Gamma_{0}$, then the phonon
excitation is thermalized and becomes incoherent. This case corresponds to the
incoherent model when Abragam theory is applicable. In the steady state
excitation, energy flowing into the ultrasonic mode from the external
perturbation equals the energy dissipated due to anharmonic coupling with the
other normal modes of the lattice causing also the fluctuations of the energy
of the excited ultrasonic mode. In quantum mechanical description, the
stronger the coupling with the other modes, the faster the decay rate of the
excited mode is.

\section{The arguments for a revision of the incoherent model}

Actually the spectra of powder absorbers \cite{Cranshaw} and thin films, for
example, stainless steel foil, experiencing mechanical vibrations,
\cite{Mishroy,Walker,Mkrtchyan77,Mkrtchyan79,Tsankov1981} are quite different.
Usually these absorbers are glued on the surface of the transducer, fed by the
oscillating voltage. The transducers, made from piezo-crystal (for example,
quartz)
\cite{Ruby,Cranshaw,Mishroy,Kornfeld,Walker,Mkrtchyan77,Perlow,Monahan,
Mkrtchyan79,Shvydko89,Shvydko92} or piezo-polymer-film (for example, a
polyvinylidene fluoride - PVDF), also produce different spectra since the
conversion factor of the PVDF drive is more than ten times larger than that of
quartz \cite{Helisto1986}. Meanwhile, the Rayleigh distribution has only one
parameter, which is variance of the displacement amplitude $\overline{a^{2}}$,
specifying also the values $\left\langle m^{2}\right\rangle $ and
$\overline{m}$. However, in general the distribution of amplitudes and phases
of the nuclear vibrations should depend on the construction of the
absorber-transducer. Therefore, it is hard to expect that by one model with a
single parameter it would be possible to fit qualitatively different
experimental spectra.

In addition to the arguments given above, the incoherent model contradicts the
observations of time domain spectra, which are obtained for $\gamma$ rays from
the vibrated source by filtering trough a single line absorber
\cite{Perlow,Monahan}. Similar experiments were performed with the vibrated
absorber and the source moved only by M\"{o}ssbauer drive
\cite{Vagizov,Shakhmuratov15,Vagizov2015,Shakhmuratov16}. In Ref.
\cite{Monahan} Monahan and Perlow developed a theory of quantum beats of
recoil-free $\gamma$ radiation, which is emitted by frequency-modulated source
and transmitted through a resonant absorber. It follows from their model that
if random phase $\psi$ and amplitude $a_{z}$ of the mechanical vibrations are
statistically independent and $\psi$ is randomly distributed over the interval
$0$ and $2\pi$, no quantum beats will be observed. If $\psi$ is distributed
normally about $\psi=0$ with variance $\left\langle \psi^{2}\right\rangle $,
then amplitudes of the harmonics in time domain spectra significantly decrease
with increase of the number of the harmonic. Since quantum beats of frequency
modulated $\gamma$-rays, which are transmitted through the resonant absorber,
are reliably observed \cite{Perlow,Monahan}, the phase $\psi$ is not randomly
distributed over the interval $0$ and $2\pi$. Moreover, no extra damping of
the second harmonic with respect to the first harmonic in the harmonic
composition of the time-dependent counting rate of the filtered $\gamma
$-photons was reported in Ref. \cite{Perlow,Monahan}. Thus, even if the
vibration phase $\psi$ is random, the phase variance $\left\langle \psi
^{2}\right\rangle $ is negligibly small.

We can add to the arguments of Perlow and Monahan that if the phase is random,
then not only the amplitudes of quantum beats of the vibrational sidebands
[see Eq. (\ref{Eq2})], observed in time domain spectra \cite{Perlow,Monahan},
are reduced or even could vanish due to the phase fluctuation of the
vibrations, but also frequency domain spectra must be broadened. This can be
shown if we consider the radiation field $E_{S}(t-t_{0})$ with vibrational
sidebands, described by Eq. (\ref{Eq2}). It is natural to suppose that the
phase $\psi$ and modulation index $m=2\pi a_{z}/\lambda$, which is
proportional to the vibration amplitude $a_{z}$, are statistically
independent. Therefore, the averaging over these parameters can be made
independently and the contribution of the amplitude and phase fluctuations are
factorized. Below we consider the contribution of the phase fluctuations.

It is well known in quantum optics that if the phase $\psi$ of the coherent
field%
\begin{equation}
E_{n}(t)=E_{c}e^{-i\omega_{S}t+in(\Omega t+\psi)}\label{Eq18}%
\end{equation}
randomly fluctuates in time, then the spectrum of the field is broadened, see
for example, Ref. \cite{Shakhmuratov98}.

Suppose that phase fluctuation follows phase diffusion process when phase
changes by small jumps and due to a random walk the phase $\psi$ can go very
far from its initial value taking all values between $0$ and $2\pi$. In this
model it is assumed that the next value of phase has a Gaussian distribution,
which is symmetric around the prior value with variance $\left\langle
\delta\psi^{2}\right\rangle $. Phase diffusion process produces spectral
broadening of the field $E_{n}(t)$. If, for example, without phase fluctuation
the field spectrum was delta-like, then due to the random walk of phase the
power spectrum of the field becomes Lorentzian with the width $\nu_{n}%
=n^{2}\left\langle \delta\psi^{2}\right\rangle /\tau_{0}$, where $\left\langle
\delta\psi^{2}\right\rangle $ is mean square value of the size of the phase
jump and $\tau_{0}$ is a mean dwell time between successive phase jumps
\cite{Shakhmuratov98}.

The phase diffusion model predicts that the central component of the field
(\ref{Eq2}) with $n=0$ is not spectrally broadened, while sidebands are
broadened. The broadening of the sidebands increases proportionally to $n^{2}%
$. To take this broadening into account we have to replace the halfwidth of
the spectral components of the field $\Gamma_{0}/2$ in Eqs. (\ref{Eq3}) and
(\ref{Eq4}) by $\gamma_{n}=\Gamma_{0}/2+n^{2}\left\langle \delta\psi
^{2}\right\rangle /2\tau_{0}$, where $n$ is the number of sideband. Usually,
all experimental spectra of the vibrated absorbers are fitted by the set of
Lorentzians with the same width. However, nobody reported progressive broadening
of the satellites.

There is another model assuming that phase after a jump takes any value
between $0$ and $2\pi$ with equal probability. This uncorrelated process
predicts the same broadening for all sidebands except the central component
with $n=0$. According to this model, Lorentzian broadening of the sidebands is
equal to $1/\tau_{0}$ \cite{Shakhmuratov98}.  As well, the marked difference
between the width of the central line and sidebands has not been yet reported.

Thus, we conclude that fast time variation of the phase of the mechanical
vibration is not present in the vibrated absorber or source if sidebands with
the same width as the central line are observed.

\section{The model of coherent vibrations with nonzero average amplitude}

The Rayleigh distribution is derived with the assumption that the amplitudes
$a_{x}$ and $a_{y}$ are Gaussian-distributed, centered at zero, and
independent, see Sec. II. This means that $a_{x,y}=0$ has maximum probability.
However, if we move a thin absorber by a coherently vibrated transducer, it is
better to suppose that the distribution of the vibration amplitudes is
centered at some value $a_{0}\neq0$. Below, following the arguments given in
Sec. III, we assume that displacements along $\mathbf{z}$ direction are
described by $a_{z}=a(x,y)\cos\Omega t$, where for simplicity we set $\psi=0$.
Then, it is natural to suppose that the amplitude $a(x,y)$ is
Gaussian-distributed and centered at $a_{0}$ with variance $\left\langle
\delta a^{2}\right\rangle $, i.e.,%
\begin{equation}
G\left(  a,a_{0}\right)  =\frac{\exp\left(  -\frac{(a-a_{0})^{2}%
}{2\left\langle \delta a^{2}\right\rangle }\right)  }{\sqrt{2\pi\left\langle
\delta a^{2}\right\rangle }}. \label{Eq19}%
\end{equation}
Here, for briefness of notations, we omit the spatial dependence of the
amplitude, i.e., $a(x,y)=a$. Usually Gaussian distribution is defined for the
value $a$ varied in the infinite domain $(-\infty,\infty)$. However, $a$ is the
amplitude, which is defined for positive values. Therefore, we restrict domain
of the amplitude variation by positive values $(0,\infty)$. To keep the same
overall density of our distribution we normalize it to the value%
\begin{equation}
N\left(  a_{0}\right)  =\int_{0}^{+\infty}G\left(  a,a_{0}\right)  da
\label{Eq20}%
\end{equation}
and obtain%
\begin{equation}
G_{\text{norm}}\left(  a,a_{0}\right)  =\frac{\exp\left(  -\frac{(a-a_{0}%
)^{2}}{2\left\langle \delta a^{2}\right\rangle }\right)  }{N\left(
a_{0}\right)  \sqrt{2\pi\left\langle \delta a^{2}\right\rangle }}.
\label{Eq21}%
\end{equation}
This distribution needs further modification since the intensity of the $n$th
sideband%
\begin{equation}
W_{n}(a_{0})=\int_{0}^{\infty}G_{\text{norm}}\left(  a,a_{0}\right)
J_{n}\left(  \frac{2\pi a}{\lambda}\right)  da, \label{Eq22}%
\end{equation}
is not zero for $n\neq0$ if $a_{0}=0$ and $\left\langle \delta a^{2}%
\right\rangle \neq0$. This discrepancy appears due to the fact that the
probability $G_{\text{norm}}\left(  a,a_{0}\right)  $ does not become zero if
$a_{0}$ is zero, i.e., when no oscillations should be present. The origin of
this discrepancy comes from the variance $\left\langle \delta a^{2}%
\right\rangle $, which should be zero if $a_{0}=0$. To fix this problem we
define variance as $\left\langle \delta a^{2}\right\rangle =(\sigma a_{0}%
)^{2}$, which means that variance is specified in a percentage $\sigma$ of the
mean value of the amplitude $a_{0}$. Then if $a_{0}=0$, then variance is also
zero. With these modifications we obtain the following expression for the
intensity of the $n$th sideband%
\begin{equation}
\overline{W}_{n}\left(  m_{0},\sigma\right)  =\frac{\sqrt{\frac{2}{\pi}}%
\int_{0}^{\infty}\exp\left[  -\frac{1}{2}\left(  x-\frac{1}{\sigma}\right)
^{2}\right]  J_{n}^{2}(\sigma m_{0}x)dx}{1+\operatorname{erf}\left(  \frac
{1}{\sqrt{2}\sigma}\right)  }, \label{Eq23}%
\end{equation}
where $m_{0}=2\pi a_{0}/\lambda$.

If $\sigma=0.1$ the variance $\left\langle \delta a^{2}\right\rangle $ is much
smaller than $a_{0}^{2}$. Then, the distribution $G_{\text{norm}}\left(
a,a_{0}\right)  $ is close to a delta-like [see Fig. 2(a)], and the dependence
of the intensity $\overline{W}_{n}\left(  m_{0},\sigma\right)  $ on $m_{0}$
[see Fig. 3(a)] is very similar to that shown in Fig. 1(a) for the coherent
model. If $\sigma=0.72$, the variance $\left\langle \delta a^{2}\right\rangle
$ is comparable with $a_{0}^{2}$. Then, the distribution $G_{\text{norm}%
}\left(  a,a_{0}\right)  $ is close to the Rayleigh distribution [see Fig.
2(b)], and the intensity $\overline{W}_{n}\left(  m_{0},\sigma\right)  $
depends on $m_{0}$ [see Fig. 3(a)] similar to that shown in Fig. 1(b) for the
incoherent model. \begin{figure}[ptb]
\resizebox{0.4\textwidth}{!}{\includegraphics{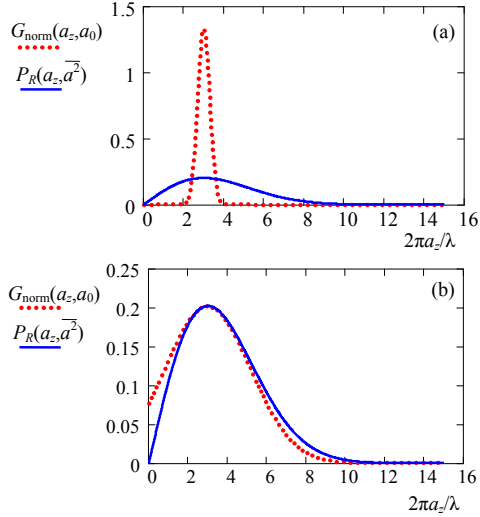}}\caption{(color on
line) The dependence of our distributions $G_{\text{norm}}\left(
a,a_{0}\right) $ (dotted line in red) and Raylegh distribution (solid line in
blue) on the amplitude of the displacement $a_{z}$. In our distribution variable
$a$ is expressed as $a_{z}$, which have the same meaning as $a$ in Eq. (\ref{Eq21}).
In both plots $a_{0}=\sqrt{\overline{a^{2}}}=3\lambda/2\pi$, which corresponds to $m_{0}=3$. Parameter $\sigma$ is 0.1 in (a), and 0.72 in (b).}%
\label{fig:2}%
\end{figure}
\begin{figure}[ptbptb]
\resizebox{0.4\textwidth}{!}{\includegraphics{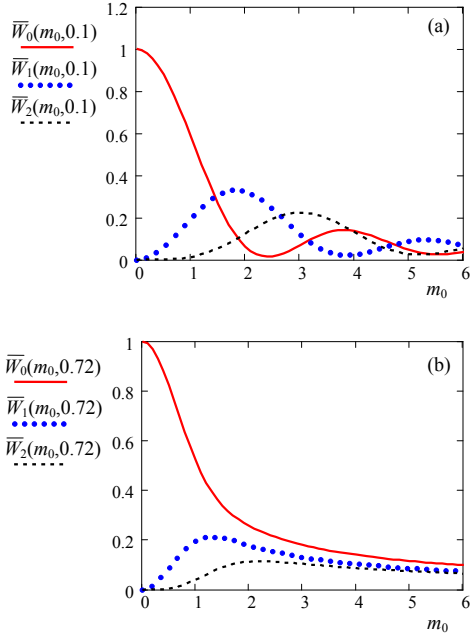}}\caption{(color on
line) Dependence of the intensity $\overline{W}_{n}\left(  m_{0}%
,\sigma\right)  $ on $m_{0}$ for our model. Parameter $\sigma$ is 0.1 in (a)
and 0.72 in (b). Solid line (in red) corresponds to $n=0$, dotted line (in blue)
shows the dependence for $n=1$, and dashed line (in black) corresponds to $n=3$.}%
\label{fig:3}%
\end{figure}

\section{Experiment}

Our experimental setup is based on an ordinary M\"{o}ssbauer spectrometer. The
source, $^{57}$Co:Rh, is mounted on the holder of the M\"{o}ssbauer drive,
which is used to Doppler-shift the frequency of the radiation of the source.

We carried out experiments with two differen absorbers. The first absorber was
made of enriched K$_{4}$Fe(CN)$_{6}\cdot3$H$_{2}$O powder with effective
thickness 13.2. It was mechanically pressed to the surface of the transducer.
Therefore, in the experiments with powder the source was mounted above the
absorber and the detector was placed below the absorber. This vertical
geometry of the experiment allowed to consider a powder as a grained substance
just freely jumping up and down under the influence of the vibrating transducer.

As a transducer we used in both experiments a polyvinylidene fluoride (PVDF)
piezo polymer film (thickness 28 $\mu$m, model LDT0-28K, Measurement
Specialties, Inc.). Several piezoelectric transducer constructions were tested
to achieve the best performance. The best of them was a piece of 28 $\mu$m
thick, 10$\times$12 mm polar PVDF film coupled to a plexiglas backing of $\sim$2
mm thickness with epoxy glue. The PVDF film transforms the sinusoidal signal
from the radio-frequency (RF) generator into a uniform vibration of the
absorber nuclei.

The second absorber was 25-$\mu$m-thick stainless-steel foil (from Alfa Aesar)
with a natural abundance ($\sim$2.2\%) of $^{57}$Fe. Optical depth of the
absorber is $T_{A}=5.18$. The stainless-steel foil is glued on the PVDF
piezotransducer. Therefore, the experiments with stainless-steel foil were
carried out in a standard horizontal geometry.

\subsection{Powder absorber}

Initially we supposed that powder absorber would behave as a sand placed on
the vibrated transducer. Then, powder grains should randomly jump and fall
down to the vibrated surface with phase depending on the size and weight of
the grains. Therefore, we expected that single parent line should not split in
sidebands, which must be strongly broadened due to the random motion of
grains, and hence vibrational sidebands could give only a broadening of the
wings of the absorption line. However, in a wide range of the vibration
frequencies from 5 MHz to 45 MHz we observed the sidebands. We used the
same voltage, 10V, supplied from RF generator, except for high frequencies (35
, 40, and 45 MHz). For them we elevated voltage up to 16V since the amplitudes
of the sidebands significantly reduced with increase of the RF frequency and
we could observe for high frequencies only first two sidebands $\omega_{S}%
\pm\Omega$ with very small intensities. \begin{figure}[ptb]
\resizebox{1\textwidth}{!}{\includegraphics{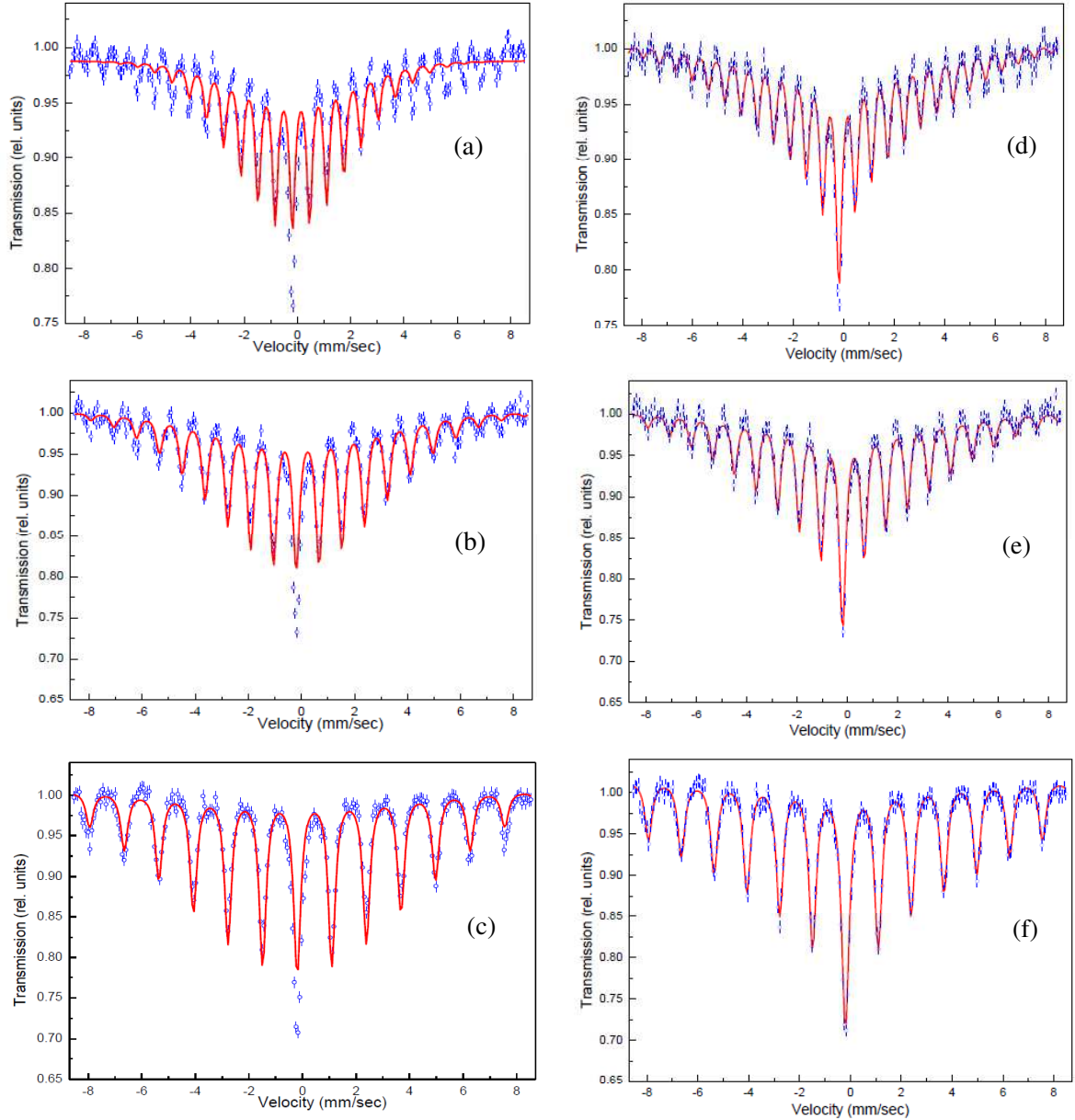}}\caption{(color on line)
Adsorption spectra of the powder absorber vibrated with frequency $\Omega$
equal to 7.5 MHz (a) and (d), 10 MHz (b) and (e), and 15 MHz (c) and (f). Dots
are experimental data, solid line (in red) is the theoretical fitting to the
Abragam model (a)-(c) and our model (d)-(f). For the Abragam model the
modulation index $m_{0}=2\pi \sqrt{\overline{a^{2}}}/\lambda$ is 3.42 in (a),
3.66 in (b), and 3.39 in (c). For our model we obtained $m_{0}$, which is 5.6
in (d), 4.3 in (e), and 4.7 in (f).}
\label{fig:4}%
\end{figure}

The fitting of the experimental data to the theoretical models is shown
in Fig. 4, where in (a)-(c) the Abragam model is used, while in (d)-(f) our model
is applied to fit data. Abragam model gives bad fitting except frequencies
$\Omega$ equal or higher than 35 MHz when we observed only three lines, i.e., the
central component and two small sidebands. For frequencies below 35 MHz, especially
the central component and components near to it are in a strong disagreement
with the Abragam model. The spectra calculated according to our model agree
well with experiment.

According to our model, for the same voltage from RF generator (10V), mean
value of the modulation index $m_{0}$ takes maximum value $m_{0}=5.6$ for
$\Omega=7.5$ MHz and then decreases with increase of the frequency $\Omega$.
For example, for $\Omega=15$ MHz we have $m_{0}=4.7$, while for  $\Omega=20$
MHz mean value of the modulation index drops almost two times, i.e., $m_{0}=2.6$.
The smallest value of the modulation index $m_{0}=0.25$ was obtained for
$\Omega=40$ MHz when the voltage was even elevated to 16V. Such a dependence
of $m_{0}$ on the modulation frequency $\Omega$ could be explained by maximum
efficiency of the process inducing mechanical vibrations of the powder
near $\Omega=7.5$ MHz.

The best fitting of the experimental data to our theoretical predictions is
obtained with $\sigma=0.85$, which corresponds to the value of the square root
of variance $\sqrt{\delta a^{2}}$ equal to 85\% of the mean amplitude $a_{0}$.
This reflects a large spread of the amplitudes of the mechanical vibrations
$a$ around $a_{0}$. Comparison of our distribution $G_{\text{norm}}(a,a_{0})$
with the Rayleigh distribution $P_{R}\left(  a_{z},\overline{a^{2}}\right)  $
for $m_{0}=5.6$ and $\sigma=0.85$ is shown in Fig. 5. Our distribution looks
close to the continuous uniform distribution for the values of $a$ between 0
and $2\lambda$. Since our powder is hygroscopic, it does not behave as a dry
sand. Such a powder can be compressed in a tablet-like substance, which shows
small adhesion to the surface of the PVDF transducer. This could explain the
coherent move of the powder grains. Their difference in size and orientation
of the crystalline axis with respect to the direction of the displacement $z$
could explain almost continuous uniform distribution of the displacement
amplitudes. \begin{figure}[ptb]
\resizebox{0.4\textwidth}{!}{\includegraphics{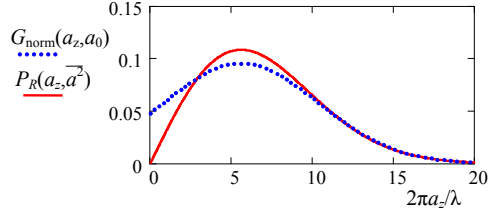}}\caption{(color on
line) The functional dependence of our distribution $G_{\text{norm}}\left(
a,a_{0}\right)  $ on the amplitude of the displacement $a_{z}$ for $m_{0}=5.6$
and $\sigma=0.85$ (dotted line in blue). The Rayleigh distribution for the
value of $\sqrt{\overline{a^{2}}}=a_{0}$, where $a_{0}=m_{0}\lambda/2\pi$, is
shown for comparison by the solid line (in red).}%
\label{fig:5}%
\end{figure}\begin{figure}[ptbptb]
\resizebox{0.5\textwidth}{!}{\includegraphics{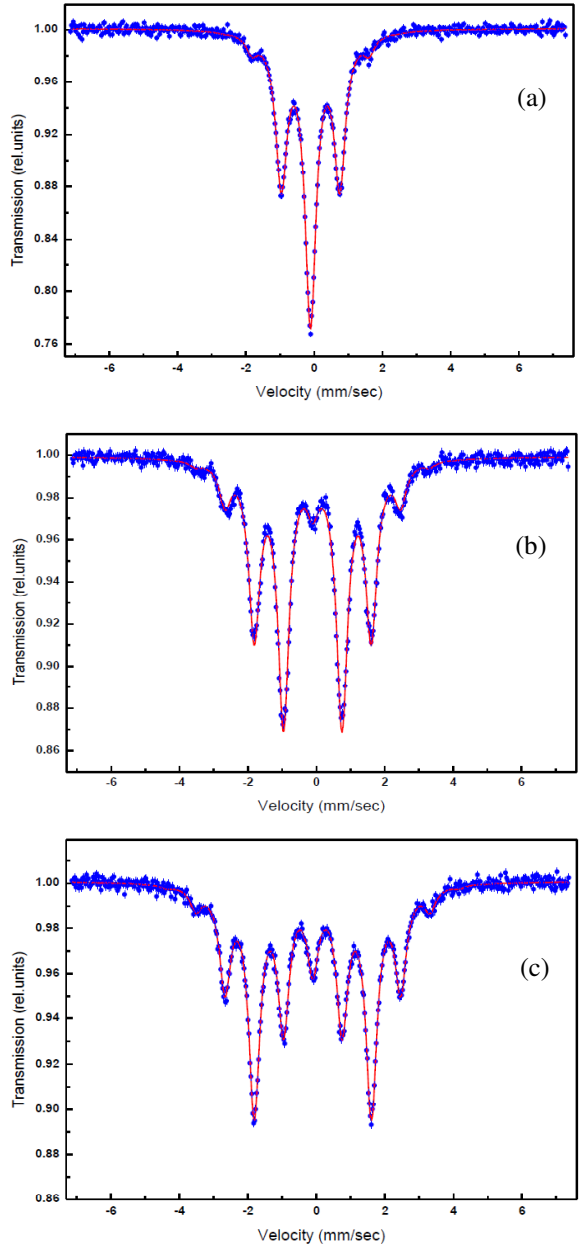}}\caption{(color on
line) Adsorption spectra for SS absorber vibrated with frequency 10 MHz. Dots
are experimental data, solid line (in red) is the theoretical fitting to our
model. Fitting parameters are $m_{0}=1.19$ and $\sigma=0.18$ in (a),
$m_{0}=2.38$ and $\sigma=0.16$ in (b), and $m_{0}=3.01$ and $\sigma=0.16$ in
(c).}%
\label{fig:6}%
\end{figure}

\subsection{Stainless steel absorber}

The spectra obtained for the stainless steel absorber are quite different from
those observed for the powder absorber. They demonstrated some features
typical for the coherent model, see Fig. 6. However, these spectra could not
be reasonably well described by the simple coherent model.

In Ref. \cite{Mkrtchyan79} it was assumed that some part of nuclei in the
absorber vibrate coherently with the same amplitude, while another part of
nuclei also take part in the coherent vibration, but for them the mean square
displacement value changes from nucleus to nucleus. We could fit experimental
data for SS absorber to the model \cite{Mkrtchyan79}, where experimental
spectra are compared with theoretical predictions assuming a mixture (a simple
sum with different weights) of the coherent and incoherent models. However,
this method did not allow to obtain good fitting. Therefore, we decided to fit
experimental spectra to our model. The results of fitting are shown in Fig. 6. \begin{figure}[ptb]
\resizebox{0.5\textwidth}{!}{\includegraphics{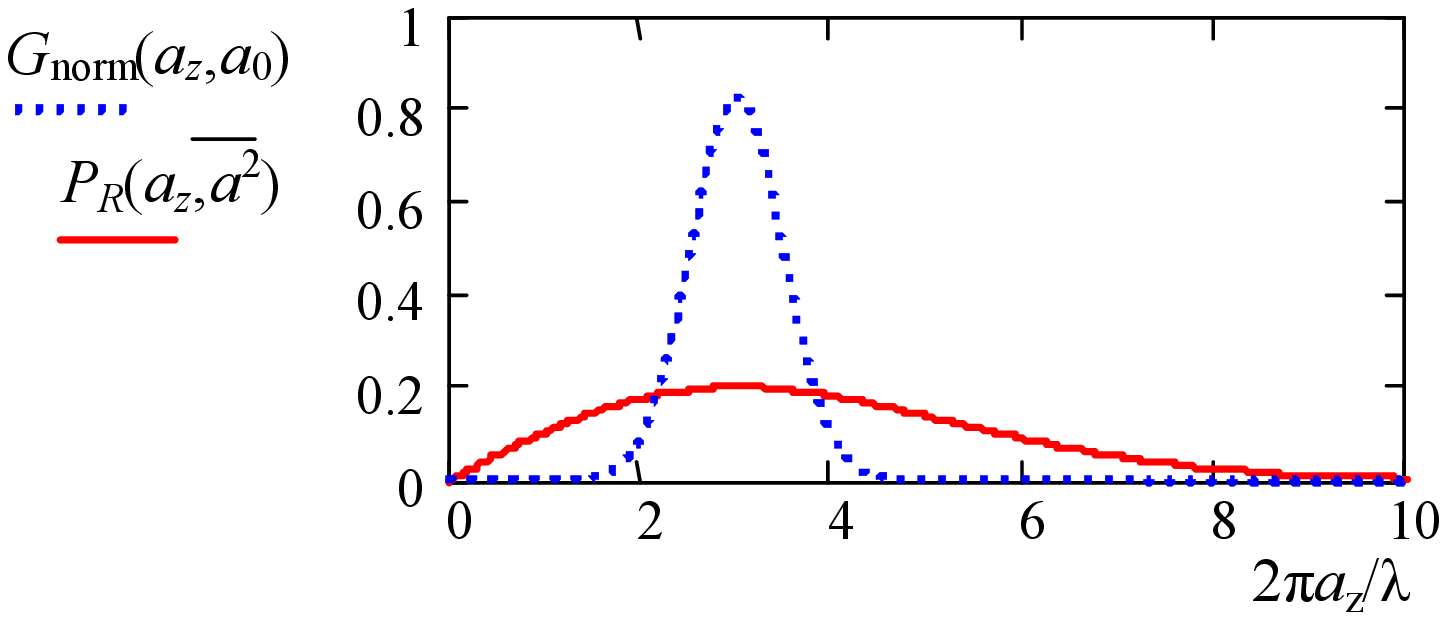}} \caption{(color on
line) Comparison of our distribution $G_{\text{norm}}\left(  a,a_{0}\right)  $
(dotted line in blue) with the Rayleigh distribution (solid line in red) for
the modulation index $m_{0}=3.01$ and $\sigma=0.16$.}%
\label{fig:7}%
\end{figure}

\subsection{Model for SS foil vibration}

The fitting allowed us to find the parameter $\sigma$. Figure 7 shows
comparison of our distribution for this parameter with the Rayleigh
distribution for the modulation index $m_{0}=3.01$. It is clear that for SS
absorber the distribution of the displacement amplitude looks bell shape. Such
a distribution gives a hint about a real distribution of the vibration
amplitudes along the surface of the absorber. Before constructing this real
distribution we give some arguments in support of the model.

PVDF film is glued on the solid plexiglas backing. Oscillating voltage forces
the film to change its thickness making it thicker or narrower. We may assume
that the lateral size of the film also oscillates. However, the solid backing
resists the lateral changes of the film size. Therefore, we may assume that
amplitude of the displacement is larger in the center of the film and smaller
at the edges.

To avoid complications we model SS film as having a form of a disc with radius
$r_{0}$. This disc vibrates such that the displacement has a maximum at the
center and decreases to the edges. All elements of the disc vibrate with the
same frequency and phase. We assume that the amplitudes of the vibrations are
distributed according to a bell-shape function, slightly resembling
$G_{\text{norm}}(a,a_{0})$, as
\begin{equation}
a_{z}(r)=a_{0}\cos\left(  \xi\frac{\pi}{2}\frac{r}{r_{0}}\right)  ,
\label{Eq24}%
\end{equation}
where $a_{0}$ is a maximum amplitude at the disc center, $r$ is a distance
from the center, and $\xi$ is a parameter, which specifies the difference
between the amplitudes at the center, $a_{z}(0)=a_{0}$, and edges,
$a_{z}(r_{0})=a_{0}\cos\left(  \xi\pi/2\right)  $. If $\xi\ll1$, then this
difference is small and the distribution is close to that when the amplitudes
are uniform along the absorber. If $\xi=1$, then the amplitude at the edges of
the disc are zero. In both cases the distribution is a bell shape as it is expected.

We assume that a beam of $\gamma$-radiation is transversely uniform and covers
the hole disk. Actually, this is not the case. In reality we have to consider
$2r_{0}$ as a diameter of the $\gamma$-beam, which is defined by the
collimator aperture and the distance from the source to absorber. Then, the
intensity of the transmitted radiation for the $n$th sideband is described by
the integral over the area $S_{0}=\pi r_{0}^{2}$
\begin{equation}
\widetilde{W}_{n}(m_{0},r_{0},\xi)=\frac{2}{r_{0}^{2}}\int_{0}^{r_{0}}%
J_{n}^{2}\left[  m_{0}\cos\left(  \xi\frac{\pi}{2}\frac{r}{r_{0}}\right)
\right]  rdr, \label{Eq25}%
\end{equation}
which is normalized to $S_{0}$, where $m_{0}=2\pi a_{0}/\lambda$. Parameter
$r_{0}$ can be excluded from the model if we introduce a variable $x=r/r_{0}$.
Then, Eq. (\ref{Eq25}) is reduced to%
\begin{equation}
\widetilde{W}_{n}(m_{0},\xi)=2\int_{0}^{1}J_{n}^{2}\left[  m_{0}\cos\left(
\xi\frac{\pi}{2}x\right)  \right]  xdx. \label{Eq26}%
\end{equation}
Thus, the parameter $\xi$ defines a measure of homogeneity of the vibration
amplitudes across the beam of $\gamma$-radiation. If $\xi\rightarrow0$, the
vibration amplitudes are almost the same for all nuclei exposed to $\gamma
$-radiation. If $\xi\rightarrow1$, the amplitudes are very different. Our
modeling assumption about the shape of the absorber is not important if
$2r_{0}$ is smaller that the lateral dimensions of the absorber. In this case
$2r_{0}$ means just the $\gamma$-beam diameter.

The dependence of the intensities of the spectral components $\widetilde
{W}_{n}(m_{0},\xi)$ for $n=0$, $1$, and $2$ is shown in Fig. 8. For small
$\xi$ ($\xi=0.1$) this dependence is close to that inherent to the coherent
model since the vibration amplitudes of nuclei are almost the same. For the
value $\xi$ ($\xi=0.95$) the dependence resembles the predictions of the
Abragam model.

The similarity of the results originates from the similarity of the structure
of the integrals in Eqs. (\ref{Eq16}) and (\ref{Eq26}) where the Bessel
function is averaged with the function proportional to $a_{z}da_{z}$ in Eq.
(\ref{Eq16}) and to $xdx$ in Eq. (\ref{Eq26}). Another common feature is that
both distributions are centered at the value of the integration variable,
which is zero, i.e., $a_{z}=0$ in Eq. (\ref{Eq16}) and $x=0$ ($r=0$) in Eq.
(\ref{Eq26}). However, these distributions are very different in one important
point. Rayleigh distribution is based on the assumption that the probability
has maximum for zero displacement $a_{z}$. Our distribution assumes that the
displacement amplitude has maximum value $\ a_{0}\neq0$ when radius $r$, which
is the averaging parameter, is zero.
\begin{figure}[ptb]
\resizebox{0.4\textwidth}{!}{\includegraphics{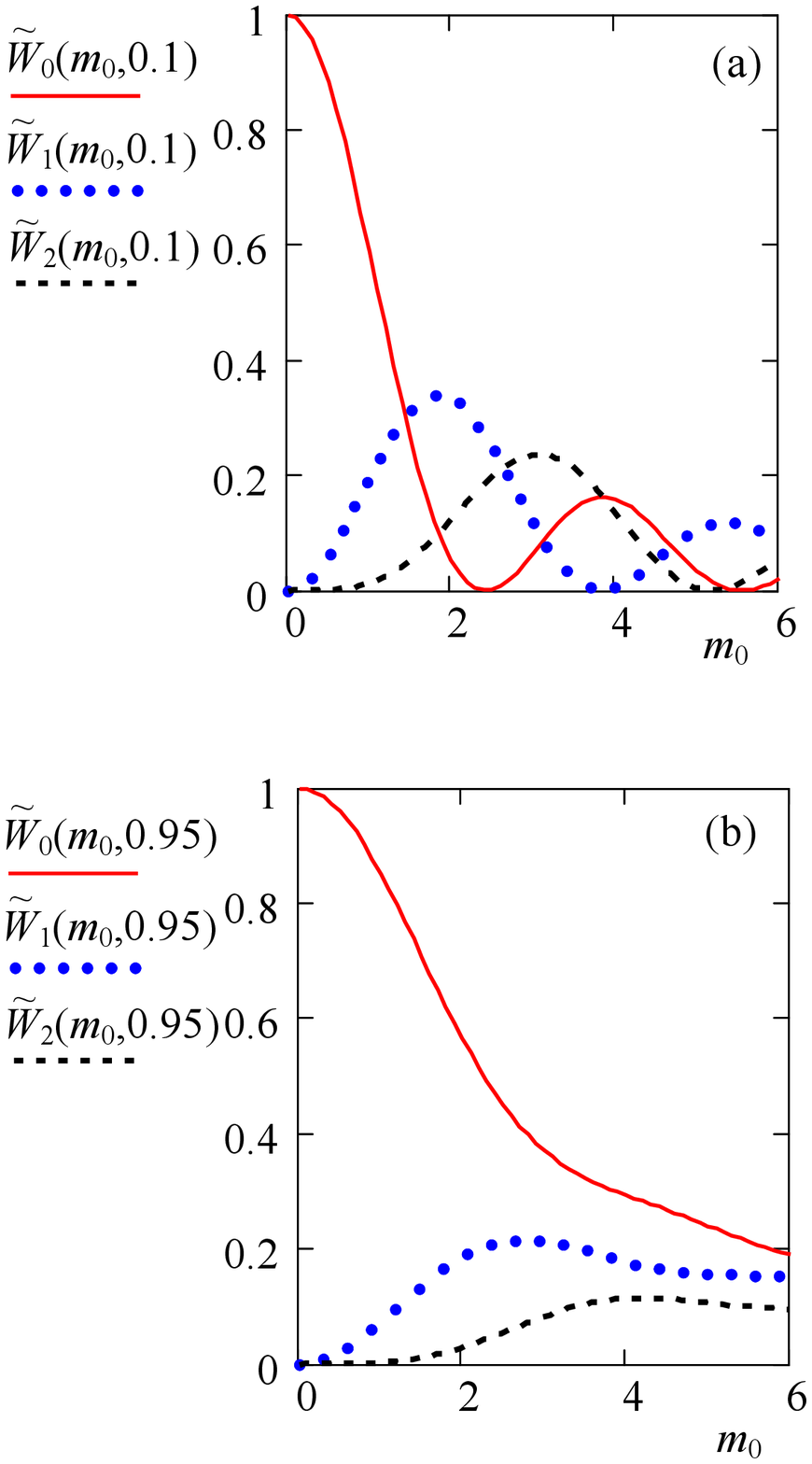}} \caption{(color on
line) Dependence of the intensity ${\widetilde{W}}_{n}\left(  m_{0}%
,\xi\right)  $ on $m_{0}$ for the model of disc vibration. Parameter $\xi$ is
0.1 in (a) and 0.95 in (b). Solid line (in red) corresponds to $n=0$, dotted line
(in blue) shows the dependence for $n=1$, and dashed line (in black) corresponds
to $n=3$.}%
\label{fig:8}%
\end{figure}

Experimental spectra are described much better by theoretical prediction,
based on the disc model, compared with our first model. Therefore, we conclude
that the disc model is more appropriate for description of the SS foil
vibration. The fitting parameter $\xi$ is quite small. This means that
dispersion of the displacements along the surface of the film is also small.

\subsection{Absorber with a mask}

We may assume that if the disc model is more adequate than other models, then
following this model it would be possible to find experimental conditions when 
the vibration amplitudes of nuclei, exposed to gamma radiation, could be made
even more homogeneous. ff we would increase the homogeneity, then the experimental 
spectra would be even more closer to those, which follow from the coherent model.

According to the disk model the simplest way to increase the homogeneity of the
displacement is to remove the contribution of nuclei located far from the absorber
center. This could be done by placing a mask with a small hole in the front of
absorber and locate the mask such that the hole coincides with the absorber center.
Then, $\gamma$-radiation will propagate only through the hole, and only nuclei,
located behind the hole will interact with $\gamma$-radiation.

To make sure that our assumptions are correct, we measured several spectra
with different diameter of the hole. According to our expectations the spectra
must change with the change of the size of the hole. \begin{figure}[ptb]
\resizebox{0.5\textwidth}{!}{\includegraphics{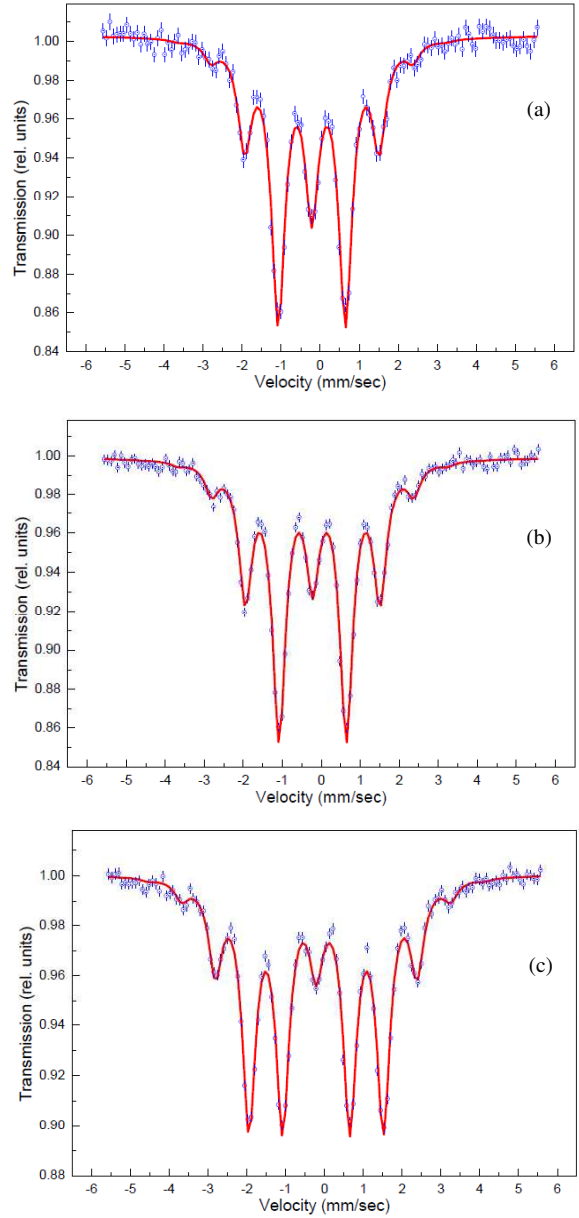}} \caption{(color on
line) Adsorption spectra for SS absorber obtained with the mask. Diameter of
the hole in the mask is 2.45 mm in (a), 1.7 mm in (b), and 1.1 mm in (c). The
absorber is vibrated with frequency 10.7 MHz. Dots are experimental data,
solid line (in red) is the theoretical fitting to the disk model. The values
of the modulation index $m_{0}$ and parameter $\xi$ are 1.85 and 0.27 in (a),
2.08 and 0.25 in (b), and 2.68 and 0.21 in (c), respectively.}%
\label{fig:9}%
\end{figure}

The absorption spectra of the absorber with the mask are shown in Fig. 9.
Diameter of the mask was varied from 2.45 mm to 1.1 mm. These spectra are
obtained for the same frequency and voltage of RF generator. In the coherent
model the central component becomes zero for the modulation index $m=2.4$. We
suppose that spectra in Fig. 9 are obtained with the modulation index quite
close to this value. Therefore, the observed lessening of the central
component of the absorption spectra with diminution of the mask diameter
proves that scattering of the vibration amplitudes of nuclei, exposed to
$\gamma$-radiation, becomes smaller with decrease of the size of the hole. At
the same time the intensities of the sidebands increase with diminution of the
hole diameter.

For the smallest hole, the disc model gives the maximum value of the vibration
amplitude $a_{0}=36.7$ pm at the disc center. The scattering parameter
$\xi=0.21$ for this hole is small. Therefore, the amplitude at the disc edge
$a_{z}(r_{0})=a_{0}\cos(\xi\pi/2)$ is 34.7 pm, which differs from $a_{0}$ only
by 2 pm. This 5$\%$ difference gives the accuracy of the displacement
measurement with the smallest hole.

By this mask with the smallest hole (1.1 mm) we scanned the surface of the
absorber. The obtained spectra are shown in Fig. 10. Since the spectra
obtained with the mask having the smallest hole are very close to that
predicted by the coherent model, we expect that such a scanning is capable to
provide the information about distribution of the vibration amplitudes along
the surface of the absorber. We obtained the following results. When the hole
coincides with the center of the absorber we have the splitting of the parent
line into sidebands, which corresponds to the modulation index $m=2.67$, see
Fig. 10(a). Positions of the hole slightly below the center and shifted to the
left in (c) and right in (d) give reduction of the modulation index to the
values $m=2.37$ in (c) and $m=2.35$ in (d). Since these values are close to
each other we may conclude that transverse shift (left/right) of the hole does
not show appreciable change of the vibration amplitude. If we move the hole
further down from the center, the value of the modulation index reduces to
$m=1.74$, see Fig. 10(b).
\begin{figure}[ptb]
\resizebox{1\textwidth}{!}{\includegraphics{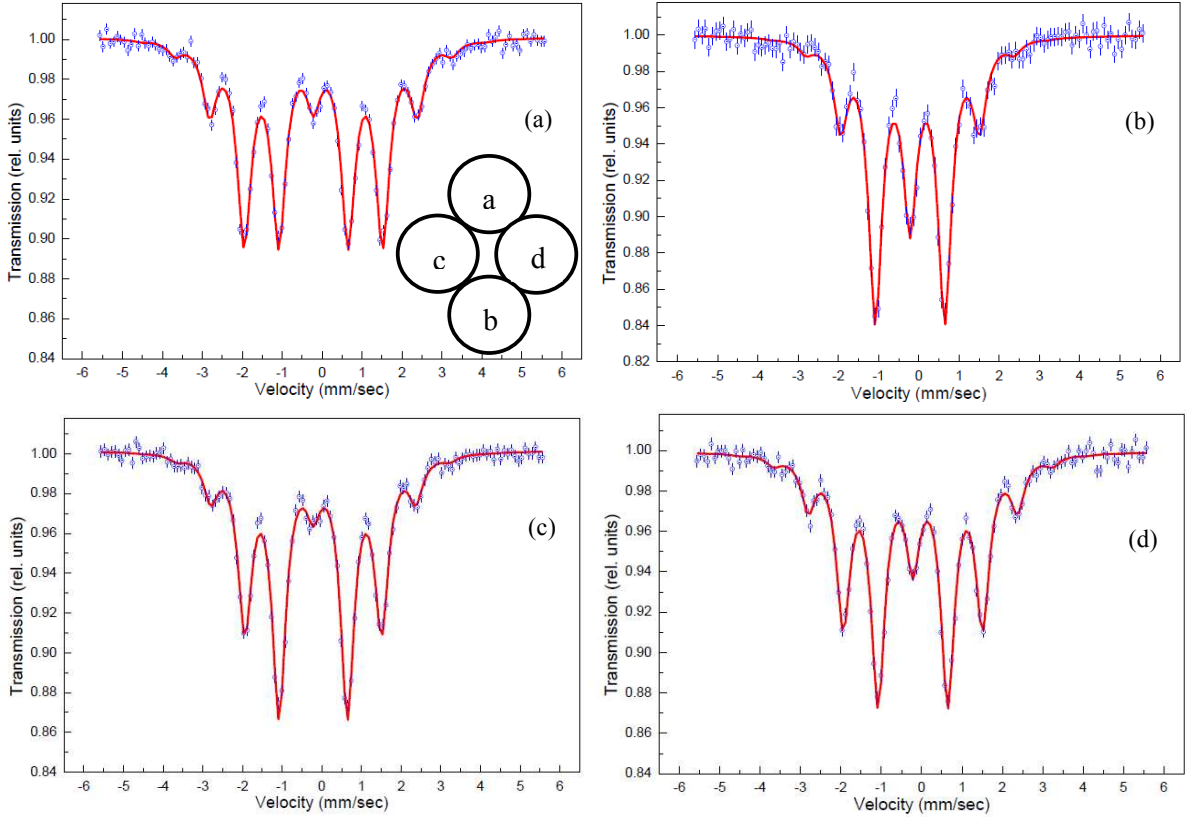}} \caption{(color on
line) Adsorption spectra for SS absorber obtained with the mask. The hole
diameter in the mask is 1.1 mm. The hole position on the absorber is shown in
the inset in (a) by black circles. Position of the upper circle coincides with
the absorber center [spectrum (a)]. Other circles are moved away to the bottom
[spectrum (b)], to the left and bottom [spectrum (c)], and to the right and
bottom [spectrum (d)]. The absorber is vibrated with frequency 10 MHz. The
driving voltage is $V=10.7$ V. Dots are experimental data, solid line (in red)
is the theoretical fitting. The fitting parameters are $m_{0}=2.67$ and
$\xi=0.18$ in (a), $m_{0}=1.74$ and $\xi=0.24$ in (b), $m_{0}=2.37$ and
$\xi=0.18$ in (c), and $m_{0}=2.35$ and $\xi=0.27$ in (b).}%
\label{fig:10}%
\end{figure}

\section{Discussion}

Our experimental results, obtained for SS foil, give a strong hint at the
presence of longitudinal distribution of the displacement amplitudes on the
surface of the absorber, while in the transverse direction the amplitudes are
more homogeneous. Since PVDF film is drawn and polarized during its
fabrication, it is natural to expect the difference of the displacements in
longitudinal and transverse directions. Long polymer chains aligned along a
particular direction give the origin to this asymmetry. Therefore, we assume
that our disc model cannot describe perfectly all the details of the vibration
of the PVDF film together with SS foil. However this model is good to describe
the experiments with the mask having a round hole. We plan to develop a strip
model of the vibration, which could be more adequate. Future experiments with
a mask, whose small hole is scanned over the surface of the absorber, could
provide topographical information about amplitude distribution over the sample
surface. We expect that this information could help to construct such a model.

As regards the powder absorber, we could screen the powder through a set of
grids to make the powder grains almost of the same size. We expect that
experiments with such a homogeneous powder could elucidate the origin of the
spectrum behavior of the vibrated powder.

\section{Conclusion}

We studied the transformation of M\"{o}ssbauer single parent line of the
vibrated absorber into a reduced intensity central line accompanied by
many sidebands. The intensities of the sidebands contain information about
the amplitudes of the mechanical vibrations and their distribution along the
surface of the absorber. Two absorbers, powder and SS foil, are experimentally
studied. The experimental spectra are fitted to the model with two parameters,
i.e., the mean amplitude of the vibrations and their deviation, which is defined
in a percentage of the mean amplitude. This model allows to conclude that the
distribution of the displacements in powder absorber is close to the continuous
uniform distribution with large scattering of the amplitudes, while for SS foil
it is bell shaped with small scattering of the amplitudes. We proposed a distribution
of the displacements in SS foil, which is related to the geometrical distribution of displacements along the surface of the foil. To verify our proposal we measured the
spectra of the vibrated SS foil placing a mask with a small hole in it before the
absorber. When the diameter of the hole in the mask is 1 mm, the displacements become
almost uniform with 5$\%$ scattering around maximum value of the displacement.
Therefore, the spectra can be described by the coherent model. This allows to
measure the displacements along the surface of the vibrated foil with the accuracy
2 pm. We expect that our finding will open a way for a new kind of spectroscopical
measurements of extra small displacements.

\section{Acknowledgements}

This work was partially funded by the Russian Foundation for Basic Research
(Grant No. 15-02-09039-a), the Program of Competitive Growth of Kazan Federal
University, funded by the Russian Government, and the RAS Presidium Program
\textquotedblleft Fundamental optical spectroscopy and its
applications.\textquotedblright

\end{document}